\documentclass[osajnl,twocolumn,showpacs,superscriptaddress,10pt]{revtex4-1} 
\usepackage{amsmath,amssymb,graphicx}
\begin{document}

\title{Photon added nonlinear coherent states for a one mode field in a Kerr medium}

\author{R. Rom\'an-Ancheyta}
\author{C. Gonz\'alez Guti\'errez}
\author{J. R\'ecamier}\email{Corresponding author: pepe@fis.unam.mx}
\affiliation{Instituto de Ciencias F\'{\i}sicas\\
 Universidad Nacional Aut\'onoma de M\'exico\\
 Apdo. Postal 48-3, Cuernavaca,
  Morelos 62251, M\'exico}

\begin{abstract}
We construct Deformed Photon Added Nonlinear Coherent States
 (DPANCSs) by application of the deformed creation operator upon the
 Nonlinear Coherent States obtained as eigenstates of the deformed
 annihilation operator and by application of a deformed displacement
 operator upon the vacuum state. We evaluate some statistical
 properties like the Mandel parameter, Husimi and Wigner functions for these states and analyze
 their differences, we give closed analytical expressions for them. We
 found a profound difference in the statistical properties of the
 DPANCSs obtained from the two above mentioned generalizations. 
\end{abstract}


\maketitle 

 \section{INTRODUCTION}
 
  The term coherent states was introduced by Glauber \cite{glauber} in
  the context of quantum optics to characterize those states of the
  electromagnetic field that factorize the field coherence function to
  all orders. The field coherent states can be constructed by any one
  of the following definitions: {\em i)} as eigenstates of the
  annihilation operator $\hat a|\alpha\rangle =\alpha|\alpha\rangle$;
  {\em ii)} as those states obtained by the application of the
  displacement operator upon the vacuum state $|\alpha\rangle =
  D(\alpha)|0\rangle$ with $D(\alpha) = \exp(\alpha \hat a^{\dagger} -
  \alpha^{*} \hat a)$ and {\em iii)} as the quantum states with a
  minimum uncertainty relationship $(\Delta p)^{2}(\Delta q)^{2} =
  (1/2)^{2}$ where the coordinate and momentum operators are defined
  as $\hat q=(1/\sqrt{2})(\hat a+\hat a^{\dagger})$, $\hat p =
  (i/\sqrt{2})(\hat a^{\dagger}-\hat a)$ with the condition $\Delta p = \Delta q
  = 1/\sqrt{2}$  \cite{glauberc,puri}.  \\ Glauber constructed the
  field coherent states by using the harmonic oscillator algebra and
  concluded that the same states are obtained from any one of the
  three mathematical definitions.  
 
 Photon added coherent states or PACS were introduced by Agarwal and
 Tara \cite{agarwal} as states that interpolate between the field
 coherent states $|\alpha\rangle$ (the most classical-like quantum
 states) and the number states $|n\rangle$ (purely quantum
 states). The unnormalized PACS are defined as  $|\alpha^{(m)}\rangle=\hat
 a^{\dagger m}|\alpha\rangle$ where the operator $\hat a^{\dagger}$ is
 the creation operator of a boson (the usual creation operator in the
 harmonic oscillator algebra) and $m$ is a positive integer that
 specifies the number of photons added to the coherent state
 $|\alpha\rangle$. These states show nonclassical features like
 squeezing, sub Poissonian statistics and negativities in their Wigner
 function due to the photons added to the coherent state. From the
 experimental point of view, these states were generated by Zavatta
 et. al. \cite{zavatta} using parametric down conversion in a
 nonlinear crystal. They reported the generation of a
 single-photon-added coherent state $|\alpha,1\rangle$ and its
 characterization by quantum tomography and reconstructed the
 corresponding Wigner function. An alternative form to generate a single photon added coherent state  was given in \cite{moya4} where, by passing a fast exited two level atom through a very high quality factor cavity which contains a coherent field, and, after the atom exits, measuring it in the ground state the system is found in a superposition of displaced number states.  It has been demonstrated that single
 photon addition and subtraction when applied to classical states of
 light produce highly nonclassical and non-Gaussian states
 \cite{zavapra72, ourjou, wenger,barbieri,lee}. More recently, Zavatta
 et. al. gave an experimental demonstration of the bosonic commutation
 relation via superpositions of quantum operations on thermal light
 fields \cite{kim,zavprl}. Single photon-added thermal states are such
 that their quasiprobability function $P(\alpha)$  does not show
 singularities and it may be accessible from experimental data
 allowing for the experimental reconstruction of a non classicality
 quasiprobability for a single photon added thermal state
 \cite{kieselpra83}. \\ 
 
It is known that the nonlinear coherent states (NLCS) constructed with
the formalism of an f-deformed algebra \cite{man'ko} also show
nonclassical properties \cite{matos,reca,dodonov}. The NLCS may be
obtained as eigenstates of a deformed annihilation operator $\hat
A|\alpha,f\rangle=\alpha|\alpha,f\rangle$, where $\hat A=\hat a f(\hat
n)$ with $f(\hat n)$ a deformation function of the number operator
$\hat n = \hat a^{\dagger} \hat a$ and also by the application of a
deformed displacement operator $\hat D_D(\alpha)=\exp(\alpha\hat
A^{\dagger}-\alpha^*\hat A)$ upon the vacuum state i.e,
$|\alpha\rangle_{D}=\hat D_D(\alpha)|0\rangle$. These two options
correspond to the generalization of Glauber's first two definitions
for the construction of a field coherent state. Recall that the
different generalizations applied to systems with dynamical properties
different from those of the harmonic oscillator yield to
non-equivalent states \cite{gilmore,reca2,roy,ren}. Sivakumar has shown
that the PACS may be regarded as NLCS for a particular deformation
function \cite{sivakumar} and in a recent publication the deformed
photon added nonlinear coherent states (DPANCSs) $|\alpha^{(m)},f\rangle$
were introduced \cite{safaeian}. These states also show nonclassical
features and for the case $f(n)=1$ reduce to the usual PACS.\\

The statistical properties of a given state may be analyzed by means
of standard parameters like the Mandel parameter, the
second order correlation function \cite{glauber63} (to test bunching
or antibunching effects), probability distributions like the Husimi
function and quasi probability distributions like the Wigner
function. The presence of  negativities in the Wigner function is a
characteristic of nonclassical states \cite{haroche,haroche2,li,li2}. 

In this work we construct DPANCSs for the case of a Kerr
Hamiltonian. We have chosen this particular system because its
spectrum contains an infinite number of bound states and its nonlinear
coherent states may be obtained exactly either as eigenstates of a
deformed annihilation operator or by application of a deformed
displacement operator upon the vacuum state \cite{octavio}.  

The paper is organized as follows: In section \ref{sec-theory}  we
briefly present the PACSs and the DPANCSs based in references
\cite{agarwal} and \cite{safaeian} respectively and in the subsection
\ref{subs-pt} we consider the specific case of a Kerr Hamiltonian and
construct the  corresponding nonlinear coherent states and their
deformed photon added nonlinear coherent states. In section
\ref{sec-estad} we analyze the statistical properties mentioned above,
that is; the Mandel parameter, Husimi Q-function, and Wigner function. Finally, in section
\ref{sec-results} we give our conclusions.

 \section{THEORY}
 \label{sec-theory}
 
Like the Nonlinear Coherent States \cite{man'ko,matos,reca} the PACS
show nonclassical properties like squeezing and sub-Poissonian
statistics. They are defined as \cite{agarwal}: 
 \begin{equation}\label{eq:pacs} 
 |\alpha^{(m)}\rangle = N_{\alpha}^{(m)} \hat a^{\dagger m}|\alpha\rangle = N_{\alpha}^{(m)}\sum_{k=0}^{m} \left(\begin{array}{c}m\\k\end{array}\right) \sqrt{k!} \alpha^{*m-k}|\alpha,k\rangle \end{equation}
with normalization constant \\ 
$N_{\alpha}^{(m)}=1/(\langle \alpha|\hat a^{m}\hat a^{\dagger m}|\alpha\rangle)^{1/2} = 1/(L_m(-|\alpha|^2)m!)^{1/2}$ \\
where $L_m(x)$ is the Laguerre polynomial of order $m$ and where $|\alpha,k\rangle= D(\alpha)|k\rangle$ is a displaced number state. We have changed the notation given in Ref.~\cite{agarwal} to avoid confusion between the PACS and the displaced number states.\medskip\\
The state $|\alpha^{(m)}\rangle$ can be written in terms of Fock states  as:
\begin{equation}
|\alpha^{(m)}\rangle=\frac{\exp(-|\alpha|^2/2)}{[L_m(-|\alpha|^2)m!]^{1/2}}
\sum_{n=0}^{\infty}\frac{\alpha^n\sqrt{(n+m)!}}{n!}|n+m\rangle.
\end{equation}

 If instead of using a coherent state $|\alpha\rangle$ one uses a
 nonlinear coherent state, for instance one constructed as eigenstate
 of a deformed annihilation operator $\hat A = \hat a f(\hat n)$ then
 the DPANCSs  (Deformed Photon Added Non Linear Coherent States) would
 be defined as \cite{safaeian}: 
 \begin{equation}
 |\alpha^{(m)},f\rangle = \frac{\hat A^{\dagger m}
   |\alpha,f\rangle}{(\langle \alpha,f|\hat A^{m}\hat A^{\dagger
     m}|\alpha,f\rangle)^{1/2}} 
 \end{equation}
 where the nonlinear coherent state $|\alpha,f\rangle$ is given by \cite{man'ko}:
 \begin{equation}\label{eq:aocs}
 |\alpha,f\rangle = N_f \sum_{n=0}^{\infty}\frac{\alpha^{n}}{\sqrt{n!} f(n)!}|n\rangle.
 \end{equation}
Here,  $N_f$ is a normalization constant,  $f(n)!=f(n)f(n-1)\cdots f(0)$ and $[f(0)]!=1$. 
 Successive applications of the deformed creation operator upon the nonlinear coherent state  $|\alpha,f\rangle$ yield: 
 \begin{equation}
 \hat A^{\dagger m}|\alpha,f\rangle = N_f \sum_{n=0}^{\infty}
 \frac{\alpha^{n} \sqrt{(n+m)!}}{n![f(n)!]^{2}}f(n+m)!|n+m\rangle.
 \end{equation}
 so that the DPANCSs can be written as:
 \begin{eqnarray}\label{eq:dpancs}
 |\alpha^{(m)},f\rangle &=&\frac{N_f}{\langle \alpha,f|\hat A^{m}\hat
   A^{\dagger m}|\alpha,f\rangle^{1/2}}\sum_{n=0}^{\infty}
\frac{\alpha^{n}\sqrt{(n+m)!}}{n!}\nonumber\\
&\times&\frac{[f(n+m)]!}{[f(n)!]^{2}}|n+m\rangle. 
 \end{eqnarray}
 
 It is clear that the states given in (\ref{eq:aocs}) and
 (\ref{eq:dpancs}) depend upon the specific form of the deformation
 function. In some cases, like for instance the Morse potential,  the
 number of bound states supported by the potential is finite and the
 states obtained from Eqs.~\ref{eq:aocs} and \ref{eq:dpancs} are
 approximate \cite{osantos2}. 
 \subsection{Anharmonic oscillator Hamiltonian}
 \label{subs-pt}
Let us consider a Hamiltonian of the form \cite{rok}
\begin{equation}
H_{D} = \hbar \Omega \hat{A}^{\dagger}\hat{A}
\end{equation}
that has been written in terms of deformed creation
$\hat{A}^{\dagger}$ and annihilation $\hat{A}$ operators defined as
$\hat{A}=\hat{a}f(\hat{n})$ and
$\hat{A}^{\dagger}=f(\hat{n})\hat{a}^{\dagger}$ where the operators
$\hat{a}$, $\hat{a}^{\dagger}$, $\hat{n}=\hat{a}^{\dagger}\hat{a}$ are
the usual harmonic oscillator operators and the function $f(\hat{n})$
is a real function of the number operator. The Hamiltonian given above
written in terms of the number operator takes the form 
\begin{equation}\label{eq:h0}
H_{D} = \hbar \Omega \hat{n} f^{2}(\hat{n}),
\end{equation}
if the function $f(\hat{n})=1$ we recover the harmonic oscillator
Hamiltonian otherwise we have a {\em deformed} oscillator with
deformation function $f(\hat{n})$. 

If we now choose the deformation function as:
\begin{equation}
f^{2}(\hat{n})= 1 +\frac{\chi}{\Omega}\hat{n},
\end{equation}
 the Hamiltonian given by Eq.~\ref{eq:h0} yields
 \begin{equation}\label{eq:kerr}
 H_{D} = \hbar\Omega\hat{n} + \hbar\chi\hat{n}^{2}\equiv
 \hbar\omega_0\left[(1-\chi/\omega_0)\ \hat{n} + (\chi/\omega_0)
   \ \hat{n}^2\right]
 \end{equation}
 which is the Hamiltonian  for a single mode field of frequency
 $\omega_0$ in a Kerr medium with a real anharmonicity parameter
 $\chi$ related to the optical properties of the
 medium\cite{mandelwolf,drummond,sanders}. It has been shown by Yurke
 and Stoler \cite{yurke} that an initial coherent state
 $|\alpha\rangle$ will evolve under the influence of an anharmonic
 oscillator Hamiltonian of the form  
 \begin{equation}\label{eq:yurke}
 H = \omega\hat{n} +\chi\hat{n}^k
 \end{equation}
into a coherent superposition of a finite number of coherent states
which are distinguishable when $|\alpha|$ is large. Notice that
Eq.\ref{eq:yurke} reduces to Eq.~\ref{eq:kerr} for $k=2$.  

 Once we have specified the deformation function the NLCS can be
 constructed explicitly using Eq.~\ref{eq:aocs}. For the case under
 consideration we obtain \cite{roman}: 
\begin{equation}\label{eq:aocspt}
|\alpha,f\rangle = N_f(\alpha) \sum_{n=0}^{\infty}
\left(\frac{\omega_0-\chi}{\chi}\right)^{n/2}\frac{\alpha^n}{\sqrt{n!\left(\omega_0/\chi\right)_{n}}}|n\rangle  
\end{equation} 
 where $\left(a\right)_{n}=\Gamma(a+n)/\Gamma(a)$ is the Pochhammer symbol and the normalization constant is
 given by
 $N_f(\alpha) = 1/\sqrt{_0F_1(\omega_0/\chi;((\omega_0-\chi)/\chi) |\alpha|^2)}$.
 Sucessive applications of the deformed creation operator upon the nonlinear coherent state $|\alpha,f\rangle$
 yield the deformed photon added nonlinear coherent state:
 \begin{eqnarray}\label{eq:simplif}
 |\alpha^{(m)},f\rangle&=&\left(N_{f}^{m}(\alpha)\right)^{-1/2}
 \sum_{n=0}^{\infty}\alpha^n
 \left(\frac{\omega_0-\chi}{\chi}\right)^{n/2}\nonumber\\
  &\times&\frac{\sqrt{(m+1)_{n}(\omega_0/\chi+m)_{n}}}
  {n!\left(\omega_0/\chi\right)_{n}}|n+m\rangle.
  \end{eqnarray}
 with the normalization constant 
$N_{f}^{m}(\alpha) = {_{2}}F_{3}(\omega_0/\chi+m,m+1;\omega_0/\chi,\omega_0/\chi,1;|\alpha|^2 (\omega_0-\chi)/\chi).$

 For a $m$ deformed photon added nonlinear coherent state
 $|\alpha^{(m)},f\rangle$, the probability of finding the $k'th$ excited
 state in the distribution is given by $P_{k,m}(\alpha) = |\langle k
 |\alpha^{(m)},f\rangle|^2$,  we obtain: 
 \begin{eqnarray}\label{prob1}
   P_{k,m}(\alpha)&=&
   |N_{f}^{m}(\alpha)|^{-1}\left(\frac{\omega_0-\chi}{\chi}\right)^{k-m}
   k!(\omega_0/\chi)_k\nonumber\\&\times&
   \frac{|\alpha|^{2(k-m)}}{m!(\omega_0/\chi)_m[(k-m)!]^2[(\omega_0/\chi)_{k-m}]^2}. 
   \end{eqnarray}
 Notice that due to the factorial in the denominator states with $k<m$
 are not allowed as was the case for the $m$ photon added coherent
 state \cite{agarwal}. 
 
\begin{figure}[h!]
\begin{center}
\includegraphics[width=8cm, height=5cm]{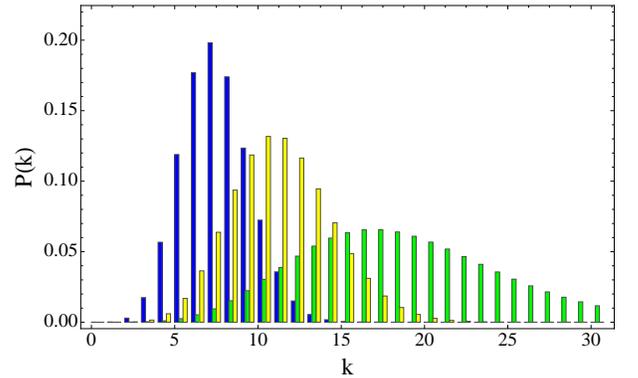} 
\caption{(Color online) Probability distributions $P_{k,m}$ with $m=1$ and
  $\alpha=3$ for a coherent state (yellow); for a NLCS obtained as
  eigenstate of the deformed annihilation operator (blue); and for a
  NLCS obtained by the deformed displacement operator acting upon the
  vacuum state (green).} 
\label{fig1}
\end{center}
\end{figure}

If we construct a deformed displacement operator by the replacement of
the usual operators $\hat a$, $\hat a^{\dagger}$ by their deformed
counterparts $\hat A$, $\hat A^{\dagger}$, we face the problem that
the commutator between them is, in general,  not a scalar and it is
not possible to write the exponential of a sum as a product of
exponentials. However, for the case we are considering here the
commutation relations are: 
  \begin{eqnarray}
  [\hat A, \hat A^{\dagger}] = 1&+&\frac{\chi}{\omega_o-\chi}
  +\frac{2\chi}{\omega_0-\chi}\hat{n},\ \ [\hat A, \hat n]=\hat A,\nonumber\\
  &&[\hat A^{\dagger},\hat n] = -\hat A^{\dagger} 
  \end{eqnarray}
  and the set of operators $\{\hat A, \hat A^{\dagger}, \hat n, 1\}$
  forms a closed Lie algebra enabling us to write the deformed
  displacement operator in a product form \cite{octavio}. 
  If we now apply the deformed displacement operator to the vacuum
  state $|0\rangle$   we obtain the nonlinear coherent states: 
  \begin{equation}\label{eq:docspt}
  |\zeta(\alpha)\rangle = (1-|\zeta(\alpha)|^2)^{\omega_0/2\chi}
  \sum_{n=0}^{\infty} \sqrt{\frac{\left(\omega_0/\chi\right)_{n}}{n!}}
  \zeta(\alpha)^{n}|n\rangle 
  \end{equation}
  where
  $\zeta(\alpha)=e^{i\phi}\tanh\left(|\alpha|/\sqrt{(\omega_0-\chi)/\chi}\right)$
  and $\alpha = |\alpha| e^{i\phi}$. 
   We can now apply $m$ times the deformed creation operator upon this
   nonlinear coherent state and obtain the $m$ photon added nonlinear
   coherent state:  
   \begin{eqnarray}\label{eq:DPADOCS}
   |\zeta(\alpha),m\rangle&=& \left(N_{\zeta(\alpha)}^{m}\right)^{-1/2}
   \sum_{n=0}^{\infty} \zeta(\alpha)^{n}\nonumber\\
	&\times&\frac{\sqrt{(m+1)_{n}(\omega_0/\chi+m)_{n}}}{n!} |n+m\rangle.  
   \end{eqnarray}
   with the normalization constant $N_{\zeta(\alpha)}^{m} =
   {_{2}}F_{1}(\omega_0/\chi +m,m+1;1;|\zeta(\alpha)|^2)$. 
   For the case of a $m$-photon added nonlinear coherent state
   constructed from the states $|\zeta(\alpha)\rangle$, the
   probability of finding the $k'th$ excited state in the distribution
   is given by: 
   \begin{eqnarray}\label{prob2}
   P_{k,m}(\zeta(\alpha)) &=& |\langle k|\zeta(\alpha),m\rangle|^2
   \nonumber\\&=&\frac{1}{N_{\zeta(\alpha)}^{m}} \frac{k!
     (\omega_0/\chi)_{k} |\zeta(\alpha)|^{2(k-m)}}{m!
     (\omega_0/\chi)_{m}[(k-m)!]^2}.
   \end{eqnarray}
 In figure \ref{fig1} we show the distribution probabilities with
 $m=1$ for the states $|\alpha^{(m)}\rangle$ (yellow), $|\alpha^{(m)},f\rangle$
 (blue) and $|\zeta(\alpha),m\rangle$ (green) for $\alpha=3$,
 $\chi/\omega_0 = 0.1$.   
   It can be seen that the distribution for the states
   $|\zeta(\alpha),m\rangle$ is broader than that of the
   $|\alpha^{(m)},f\rangle$ with that for the states $|\alpha^{(m)}\rangle$ in
   between. A more detailed analysis of this conduct will be presented
   later in this work when we evaluate several statistical properties
   like the Mandel parameter, the Husimi Q-function
   and the Wigner function. 

   \section{STATISTICAL PROPERTIES}
   \label{sec-estad}
\begin{figure}[h!]
\begin{center}
\includegraphics[width=8cm, height=5cm]{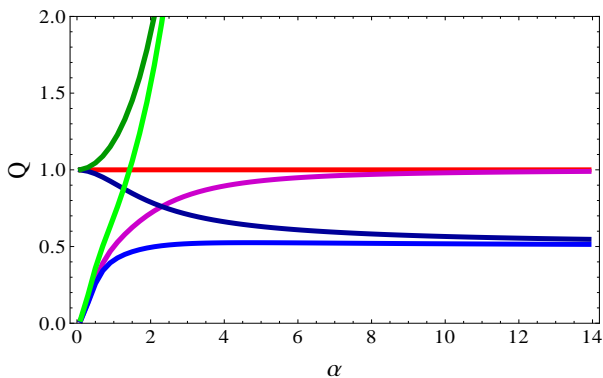} 
\caption{(Color online) Mandel parameter $Q$ with $m = 0, 1$ as a function of
  $\alpha=(x,0)$ for a m-photon added coherent state
  $|\alpha^{(m)}\rangle$ (red, purple), and deformed m-photon added non linear
  coherent states $|\alpha^{(m)},f\rangle$ (dark blue, blue) and
  $|\zeta(\alpha),m\rangle$ (dark green, green) with the parameter
  $\chi/\omega_0=0.15$.} \label{fig2} 
\end{center}
\end{figure}

   We now analyze some statistical properties of the states we constructed in the previous section. Several criteria have been developed  to analyze non-classical states \cite{ching}.     Consider for instance the Mandel parameter  \cite{mandel1} defined as: 
   \begin{equation}
   Q = \frac{\langle \hat n^2\rangle -\langle \hat n\rangle^2}{\langle \hat n \rangle},
   \end{equation}
For a Poissonian distribution the Mandel parameter is $Q=1$
corresponding to a field coherent state $|\alpha\rangle$, for $Q>1$ we
have  super Poissonian statistics (classical states) and for $Q<1$ sub
Poissonian statistics (non classical states).
 The Mandel parameter as a function of $\alpha$ using the m-photon
 added coherent states $|\alpha^{(m)}\rangle$ as well as m-photon added
 non linear coherent states $|\alpha^{(m)},f\rangle$ and
 $|\zeta(\alpha),m\rangle$ is shown in figure {\ref{fig2}} for the
 case where $\chi/\omega_0=0.15$. As mentioned above, the Mandel
 parameter for a field coherent state $|\alpha\rangle$ is a constant equal to
 one. For a photon added coherent state with $m$ different from zero
 the Mandel parameter is zero when $\alpha$ is zero and increases with $\alpha$
 tending asymptotically to one, in the figure we present the case when
 $m=1$.    

    The Mandel parameter using the nonlinear coherent states
    $|\alpha^{(m)},f\rangle$ and $|\zeta(\alpha),m\rangle$ is also shown
    in figure \ref{fig2}. For the case with $m=0$ (no photon added)
    and for small values of $\alpha$ both calculations converge to the
    harmonic result, that is, they start at one. As the $\alpha$ value
    increases their behavior separates from the harmonic one. For the
    nonlinear states $|\alpha^{(0)},f\rangle$ the Mandel parameter takes
    values smaller than one indicating a non classical behavior in
    contrast with the nonlinear states $|\zeta(\alpha),0\rangle$ whose
    Mandel parameter is larger than one indicating a super Poissonian
    statistics. For $m\neq 0$ (in the figure we present the case m=1),
    their Mandel parameter starts at zero in accordance with the PACS
    $|\alpha^{(m)}\rangle$, however as the $\alpha$ value increases, the
    Mandel parameter for the states $|\alpha^{(1)},f\rangle$ approaches
    0.6 asymptotically in contrast with the PACS  whose asymptotic
    value is that corresponding to a coherent state. That means that
    independently of the value of the parameter $\alpha$ these
    deformed photon added nonlinear coherent states remain
    nonclassical. The Mandel parameter for the non linear states
    $|\zeta(\alpha),1\rangle$ starts at zero as the other two cases
    but attains values larger than one indicating a classical behavior
    for large enough values of $\alpha$.  
    
It is worth mentioning that for the Hamiltonian given by
Eq.~\ref{eq:kerr}, both nonlinear coherent states $|\alpha,f\rangle$
and $|\zeta(\alpha)\rangle$ are exact because the number of bound
states is infinite.

    The Husimi-Q function is always positive, and in some cases is
    convenient to use for simple representations of quantum states.  
    It is defined as the Fourier transform of the anti-normal-order
    characteristic function 
    \[ Q^{[\rho]}(\alpha)= \frac{1}{\pi^2}\int d^2\lambda e^{(\alpha
      \lambda^{*} -\alpha^{*} \lambda)} C_{an}^{[\rho]}(\lambda), \] 
    which is simply related to the expectation value of the density
    operator $\rho$ in state $|\alpha\rangle$ by: 
    \begin{equation}
    Q^{[\rho]}(\alpha)= \frac{1}{\pi} {\rm Tr}\ [\rho |\alpha\rangle
      \langle \alpha|] = \frac{1}{\pi} \langle
    \alpha|\rho|\alpha\rangle 
    \end{equation}
    which can also be written as:
    \begin{eqnarray}\label{eq:husimi}
    Q^{[\rho]}(\alpha)&=&\frac{1}{\pi}\langle 0|D(-\alpha)\rho
    D(\alpha)|0\rangle\nonumber\\
	 &=& \frac{1}{\pi} {\rm Tr} \ [|0\rangle \langle
      0| D(-\alpha)\rho D(\alpha)]. 
    \end{eqnarray}
    The $Q$ function is thus the average of the projector onto the
    vacuum state, in the field displaced in phase space by
    $-\alpha$. Eq.~\ref{eq:husimi} is the basis of the Q-function reconstruction method described in Refs.~\cite{haroche,haroche2}. 
    
\begin{figure}[h!]
\begin{center}
\includegraphics[width=8cm, height=6cm]{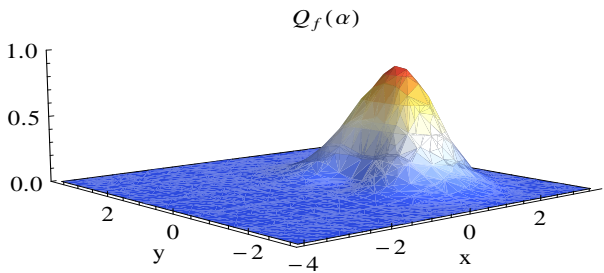} 
\caption{(Color online) Husimi Q-function $Q^{[|\alpha^{(1)},f\rangle \langle
      \alpha^{(1)},f|]}(z)$ with  $\alpha=1.1$ for a nonlinear photon added
  coherent state $|\alpha^{(1)},f\rangle$, and a coherent state
  $|z\rangle$  with $z = x + i y$ and
  $\chi/\omega_0=0.15$.} \label{husimiaocsm3} 
\end{center}
\end{figure}

    For  coherent states $|z\rangle$ it is given by:
    \begin{equation}
    Q^{[|\alpha\rangle\langle \alpha|]}(z) =\frac{1}{\pi} \langle
    z|\alpha\rangle\langle \alpha|z\rangle=\frac{1}{\pi}
    \exp{(-|z-\alpha|^2)}. 
    \end{equation}
    with $|z\rangle$ and $|\alpha\rangle$ usual harmonic oscillator coherent states.
     For the case of a coherent state $|z\rangle$ and an $m$ photon
     added coherent state $|\alpha^{(m)}\rangle$ the Husimi function is
     defined by  \cite{agarwal} : 
     \begin{eqnarray}
      Q^{[|\alpha^{(m)}\rangle\langle \alpha^{(m)}|]}(z)=\frac{1}{\pi} \langle z|\alpha^{(m)}\rangle\langle \alpha^{(m)}|z\rangle
	\nonumber\\=\frac{1}{\pi} \frac{|z|^{2m}}{m! L_m(-|\alpha|^2)}\exp{(-|z-\alpha|^2)}\ \
      \end{eqnarray}
      We now consider the deformed $m$ photon added nonlinear coherent
      state $|\alpha^{(m)},f\rangle$ given by Eq.~\ref{eq:simplif}. The
      Husimi function is in this case: 
      \begin{equation}
      Q^{[|\alpha^{(m)},f\rangle \langle \alpha^{(m)},f|]}(z)= \frac{1}{\pi} \langle
      z|\alpha^{(m)},f\rangle \langle \alpha^{(m)},f|z\rangle 
      \end{equation}
      and taking the products we get:
      \begin{eqnarray}
      &Q&^{[|\alpha^{(m)},f\rangle \langle \alpha^{(m)},f|]}(z)=\frac{1}{\pi} \frac{e^{-|z|^2}}{N_{\alpha}^{m,f}} \frac{|z|^{2m}}{m!}
\nonumber\\&&\times\left| \sum_{n=0}^{\infty}\frac{(z^{*}\alpha)^{n}}{n!}
\left(\frac{\omega_0 -\chi}{\chi}\right)^{n/2}
\frac{\sqrt{(\omega_0/\chi+m)_{n}}}{(\omega_0/\chi)_{n}}
\right|^{2}
      \end{eqnarray}
with the normalization constant $N_{\alpha}^{m,f}={_2F_3(\omega_0/\chi+m, m+1; \omega_0/\chi,
   \omega_0/\chi, 1;(\omega_0-\chi)/\chi |\alpha|^2)}$. 
    In figure \ref{husimiaocsm3} we show the Husimi function $
    Q^{[|\alpha^{(1)},f\rangle \langle \alpha^{(1)},f|]}(z)$ (1 photon added)
    for  $\alpha=1.1$, $z= x + i y$, $\chi/\omega_0 = 0.15$.

    For the case of deformed photon added nonlinear coherent states
    obtained by application of the deformed displacement operator upon
    the vacuum state $|\zeta(\alpha),m\rangle$ (see
    Eq.~\ref{eq:DPADOCS}) we obtain: 
    \begin{eqnarray}
    Q^{[|\zeta(\alpha),m\rangle \langle \zeta(\alpha),m|]}(z) =\frac{1}{\pi}
    \frac{e^{-|z|^2}}{N_{\zeta(\alpha)}^{m}}\frac{|z|^{2m}}{m!}\times\nonumber\\
    \left|\sum_{n=0}^{\infty}
    \frac{(z^{*}\zeta(\alpha))^{n}}{n!}\sqrt{(\omega_0/\chi+m)_{n}}\right|^{2}. 
    \end{eqnarray}
    with normalization constant $N_{\zeta(\alpha)}^{m}=
    {_2}F_1(\omega_0/\chi+m, m+1;1;|\zeta(\alpha)|^2)$.  
   
\begin{figure}[h!]
\begin{center}
\includegraphics[width=8cm, height=6cm]{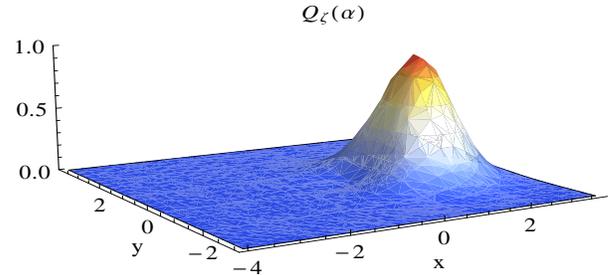} 
\caption{(Color online) Husimi Q-function $Q^{[|\zeta(\alpha),1\rangle \langle
      \zeta(\alpha),1|]}(z)$ with $\alpha = 1.1$, for a nonlinear photon
  added coherent state $|\zeta(\alpha),1\rangle$, and a coherent state
  $|z\rangle$  with $z = x + i y$ and
  $\chi/\omega_0=0.15$.} \label{husimidocsm3} 
\end{center}
\end{figure}

   In figure \ref{husimidocsm3} we show its Husimi function with the
   same set of parameters used in figure \ref{husimiaocsm3}. Notice
   that both distributions are qualitatively similar even though the
   nonlinear coherent states $|\alpha^{(m)},f\rangle$ and
   $|\zeta(\alpha),m\rangle$ differ significantly in their Mandel parameter.      
    We also calculated the Husimi function for no-photon added
    nonlinear coherent states and the differences between them are too
    small to be noticed in a graph for this value of the parameter $\alpha$.

Since the Husimi function is always positive it is best to compute the
Wigner function for the purpose of getting information related with
the non classicality  of the state \cite{li}. To that end we consider the work
of Refs.~\cite{moya,moya2,royer} where the Wigner function is written as an
infinite series in terms of the displaced number-state expectation
values
\begin{equation}
W(\alpha)= \frac{2}{\pi}\sum_{k=0}^{\infty} (-1)^{k} \langle
\alpha,k|\hat\rho|\alpha,k\rangle  
\end{equation}
and where the states $|\alpha,k\rangle$ are defined as \cite{oliveira,boiteux,moya3,moya4}:
\begin{eqnarray*}
|\alpha,k\rangle & = & \hat D(\alpha)|k\rangle \\ & = & \exp(-|\alpha|^2/2) \sum_{n=0}^{\infty} \left(\frac{k!}{n!}\right)^{1/2} \alpha^{k-n}{\cal{L}}_{n}^{k-n}(|\alpha|^2) |n\rangle,
\end{eqnarray*}
where ${\cal{L}}_{n}^{k-n}(x)$ is the associated Laguerre polynomial.
The displaced number state is a generalization of the usual field coherent state, they have been studied theoretically and experimentally.
It is worth to mention that the displaced Fock states of the electromagnetic field were synthesized by overlapping the single-photon Fock state $|1\rangle$ with coherent states on a high-reflection beam splitter and completely characterized by means of quantum homodyne tomography \cite{lvovsky} and more recently displaced number states of a vibrational mode of a single trapped ion were created by means of a combination of optical and electrical manipulation of the ion \cite{ziesel}.

The density matrix for a coherent state is
$\hat\rho = |\beta\rangle\langle \beta|$,
so that the Wigner function can be written as \cite{moya}:
\begin{eqnarray} W(\alpha) & = & \frac{2}{\pi}\sum_{k=0}^{\infty}
  (-1)^{k} \langle \alpha,k|\beta\rangle \langle \beta|\alpha,k\rangle
  \nonumber\\&=&\frac{2}{\pi}\exp{(-2|\beta-\alpha|^2)}
\end{eqnarray} 
For the construction of the Wigner function corresponding to the
nonlinear coherent state $|\beta,f\rangle$ we get: 
\begin{equation}\label{eq:wignerd}
W_f(\alpha)=\frac{2}{\pi}\sum_{k=0}^{\infty} (-1)^{k} \langle
\alpha,k|\beta,f\rangle \langle \beta,f|\alpha,k\rangle 
\end{equation}
where $|\beta,f\rangle$ is the nonlinear coherent state obtained
either as eigenstate of the deformed annihilation operator
(Eq.~\ref{eq:aocspt})  or by displacement of the vacuum state by the
generalized displacement operator (Eq.~\ref{eq:docspt}). 

Since the states $|\alpha^{(m)},f\rangle$ and $|\zeta(\alpha),m\rangle$
show different statistical behavior, we will consider them
separately. First we  take the density matrix corresponding to a
nonlinear photon added coherent state obtained as an eigenstate of the
annihilation operator. The Wigner function is: 
\begin{equation}
W_{f,A}^m(\alpha) = \frac{2}{\pi}\sum_{k=0}^{\infty} (-1)^{k} \langle
\alpha,k|\beta^{(m)},f\rangle \langle \beta^{(m)},f|\alpha,k\rangle 
\end{equation}
where the state $|\beta^{(m)},f\rangle$ is that given by
Eq.~\ref{eq:simplif}. As a result we obtain: 
\begin{eqnarray}\label{eq:wigneraocs}
W_{f,A}^m(\alpha)&=&\frac{2e^{-|\alpha|^2} }{\pi N_f^{m}(\beta)}
\sum_{k=0}^{\infty}(-1)^{k}
|\alpha|^{2(m-k)}|\sum_{l=0}^{k}\frac{(-|\alpha|^2)^l}{l!}\nonumber\\ 
&\times&\sum_{n=0}^{\infty}\frac{(\alpha^{*}\beta)^n }{n!}\frac{k!}{m!}
\left(\frac{\omega_0-\chi}{\chi}\right)^{n/2}{n+m \choose k-l}\nonumber\\
&\times&\frac{\sqrt{(\omega_0/\chi+m)_n}}{(\omega_0/\chi)_n}|^2. 
\end{eqnarray}

 The Wigner function for m-photon added nonlinear coherent states
 $|\zeta(\beta),m\rangle$ is:
\begin{equation} 
 W_{f,D}^{m}(\alpha)=\frac{2}{\pi}\sum_{k=0}^{\infty} (-1)^k\,\langle\alpha,k
 |\zeta(\beta),m\rangle \langle\zeta(\beta),m| \alpha, k\rangle
\end{equation}
with the states $|\zeta(\beta),m\rangle$ given by
Eq.~\ref{eq:DPADOCS}. As a result we get: 
\begin{eqnarray}\label{eq:wignerdocs}
W&&_{f,D}^{m}(\alpha)=	
 \frac{2e^{-|\alpha|^2}}{\pi N_{\zeta(\beta)}^{m}}
 \sum_{k=0}^{\infty}(-1)^{k}
|\alpha|^{2(m-k)}|\sum_{l=0}^{k}\frac{(-|\alpha|^2)^l}{l!}\nonumber\\ 
&&\times\sum_{n=0}^{\infty}\frac{[\alpha^{*}\zeta(\beta)]^n }{n!}\frac{k!}{m!}
{n+m \choose k-l}\sqrt{(\omega_0/\chi+m)_n}|^2.
\end{eqnarray}

When we evaluated the Mandel parameter, we found that the states
$|\zeta(\alpha)\rangle$ have a super-Poissonian statistics while
$|\alpha,f\rangle$ a sub-Poissonian one and both coincide with the
usual coherent states when $\alpha$ is small enough. This means that
nonclassical behavior is expected for  $|\alpha,f\rangle$ and not for
$|\zeta(\alpha)\rangle$. For the photon added nonlinear coherent states $|\zeta(\alpha),m\rangle$ and $|\alpha^{(m)},f\rangle$ the Mandel parameter is smaller than one for small values of the parameter $\alpha$. We should expect a nonclassical behavior for both nonlinear coherent states in this region of $\alpha$.  

It has been found experimentally that the single photon added coherent
state has a Wigner function that can be negative for small values of
the parameter $\alpha$ and the state approaches a coherent state
conduct for large enough values of it \cite{zavatta}. 

\begin{figure}[h!]
\begin{center}
\includegraphics[width=8cm, height=6cm]{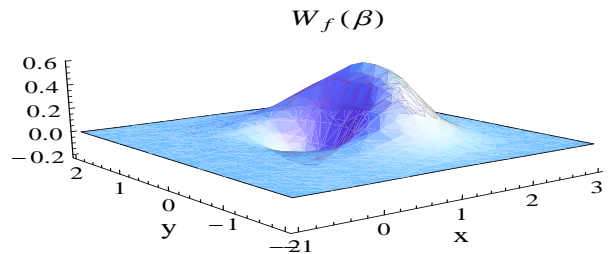} 
\caption{(Color online) Wigner function $W_{f,A}^{m=1}(\alpha)$ with a density matrix
  corresponding to a nonlinear coherent state $|\beta^{(1)},f\rangle$
  obtained as eigenstate of the annihilation operator with $\beta=1.1$,
  $\chi/\omega_0=0.15$.} \label{fig8} 
\end{center}
\end{figure}

\begin{figure}[h!]
\begin{center}
\includegraphics[width=8cm, height=6cm]{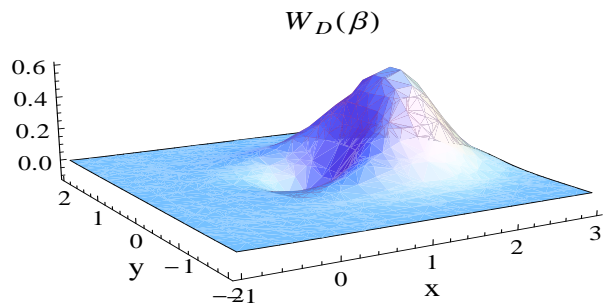} 
\caption{(Color online) Wigner function $W_{f,D}^{m=1}(\alpha)$ with a density matrix
  corresponding to a nonlinear coherent state $|\zeta(\beta),1\rangle$
  obtained by displacement of the vacuum state with $\beta=1.1$,
  $\chi/\omega_0=0.15$.} \label{fig9} 
\end{center}
\end{figure}

We evaluated the expressions given in
Eqs.~\ref{eq:wigneraocs} and \ref{eq:wignerdocs} and the results are
shown in  figures \ref{fig8} and \ref{fig9}.  

We can see from the figures that in both cases the Wigner function presents negativities and in figure \ref{fig8} these negativities are more pronounced than those in figure \ref{fig9}. This result is in agreement with what we found when we calculated the Mandel parameter, since for this value of the parameter $\alpha$ the Mandel parameter is in both cases smaller than one, though in the case of the $|\zeta(\alpha),1\rangle$ it is close to one whereas for the states $|\alpha^{(1)},f\rangle$ it is about 0.5.
 
\begin{figure}[h!]
\begin{center}
\includegraphics[width=8cm, height=6cm]{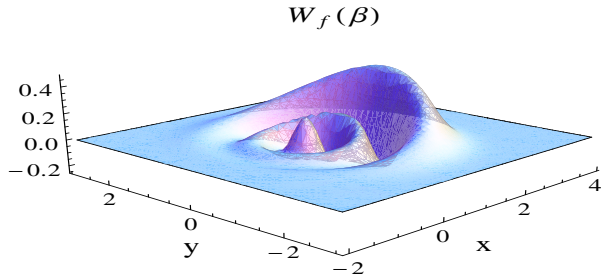} 
\caption{(Color online) Wigner function $W_{f,A}^{m}(\alpha)$ with a density matrix
  corresponding to a photon added nonlinear coherent state
  $|\beta^{(m)},f\rangle$ obtained as eigenstate of the deformed
  annihilation operator with $\beta=0.5$, $\chi/\omega_0=0.15$ and
  $m=4$.} \label{fig10} 
\end{center}
\end{figure}

 In figure
\ref{fig10} we show the Wigner function for a photon added nonlinear
coherent state $|\beta^{(m)},f\rangle$ for the case with $m=4$ and it is
seen that it takes negative values and the distribution is not
symmetrical with respect to the axes.  

\section{CONCLUSIONS}
\label{sec-results}

In this work we have analyzed the statistical differences between
nonlinear coherent states for a Kerr like medium when one makes use of
two different generalizations for their construction. On the one hand
we have generalized the definition as eigenstates of a deformed
annihilation operator (AOCS), on the other, as those states obtained by the
application of a generalized displacement operator upon the vacuum
state (DOCS). Since the Hamiltonian fulfills a finite Lie algebra and the
number of eigenvalues is infinite, both generalizations are exact and thus allow by the measurement of their statistical properties to discern which generalization is most appropriate, at least for this particular Hamiltonian. It would be difficult to obtain this information from the reconstruction of the Husimi or the Wigner functions since these are very similar for the AOCS and the DOCS.  

When computing the Mandel parameter we
found that the AOCS present a sub-Poissonian statistics corresponding
to non classical states while the DOCS present a super-Poissonian
statistics corresponding to a thermal state.   
We constructed also deformed m-photon added nonlinear coherent states
by application of the deformed creation operator upon these non linear
coherent states. We found that the deformed m-photon added nonlinear
coherent states obtained by application of the deformed annihilation
operator  upon the $|\alpha,f\rangle$ present a more pronounced non
classicality than the m-photon added coherent states introduced by
Agarwal and Tara and our results are in agreement with those of
Ref.\cite{safaeian}. However, when we construct the deformed m-photon
added nonlinear coherent states  by application of the deformed
annihilation operator  upon $|\zeta(\alpha)\rangle$ we obtain states
whose conduct becomes classical for large enough values of the
parameter $\alpha$.\\ 
We stress the fact that for a deformation function $f(\hat n)$
corresponding to a Kerr like medium the non linear coherent states
$|\alpha,f\rangle$ and $|\zeta(\alpha)\rangle$ present a completely
different statistical behavior, in the first case the states are non
classical for all values of the parameter $\alpha$, in the second 
they behave as classical states. When one constructs the deformed
photon added nonlinear coherent states with $|\alpha,f\rangle$ they
are more non classical than the PACS, while those constructed with the
$|\zeta(\alpha)\rangle$ are nonclassical only for small $\alpha$ and
become classical for large enough values of it. 
\vspace{.5cm}

\noindent{\bf Acknowledgements:} We thank Reyes Garc\'{\i}a for the
maintenance of our computers and acknowledge partial support from
CONACyT  through project 166961 and DGAPA-UNAM project IN108413.

 \end{document}